\documentstyle[12pt]{article}
\def\beq{\begin{equation}}
\def\eeq{\end{equation}}
\def\bea{\begin{eqnarray}}
\def\eea{\end{eqnarray}}
\def\bq{\begin{quote}}
\def\eq{\end{quote}}

\def\JP{{\it J.Phys.} }

\def\NP{{\it Nucl.Phys.} }
\def\PL{{\it Phys.Lett.} }
\def\PR{{\it Phys.Rev.} }
\def\PRL{{\it Phys.Rev.Lett.} }

\def\ZP{{\it Z.Phys.} }

\def\gappeq{\mathrel{\rlap {\raise.5ex\hbox{$>$}}
{\lower.5ex\hbox{$\sim$}}}}

\def\lappeq{\mathrel{\rlap{\raise.5ex\hbox{$<$}}
{\lower.5ex\hbox{$\sim$}}}}

\begin{document}
\pagestyle{empty}
\begin{flushright}
{CERN-TH/95-273}
\end{flushright}
\vspace*{5mm}
\begin{center}
{\bf THE PION STRUCTURE FUNCTION IN A CONSTITUENT MODEL} \\
\vspace*{1cm}
{\bf G.~Altarelli} \\
\vspace{0.3cm}
Theoretical Physics Division, CERN\\
CH - 1211 Geneva 23 \\
\vspace{0.5cm}
{\bf S.~Petrarca and F.~Rapuano} \\
\vspace{0.3cm}
Dip. di Fisica dell'Universit\`{a} {\it La Sapienza} \\
P.le A. Moro 2, 00185 Roma, Italy \\
and \\
INFN, Sez. di Roma I \\
P.le A. Moro 2, 00185 Roma, Italy \\
\vspace*{2cm}
\noindent
{\bf ABSTRACT} \\ \end{center}
\vspace*{5mm}
Using the recent relatively precise experimental results on the pion structure
function, obtained from Drell--Yan processes, we quantitatively test an old
model where the structure function of any hadron is determined by that of its
constituent quarks. In this model the pion structure function can be predicted
from the known nucleon structure function. We find that the data support the
model, at least as a good first approximation.

\vspace{5cm}
\noindent

\begin{flushleft} CERN-TH/95-273 \\
October 1995
\end{flushleft}
\vfill\eject

\setcounter{page}{1}
\pagestyle{plain}

\section{Introduction}

In 1974 a model was proposed \cite{acmp} for the deep inelastic scattering
structure functions of a hadron in terms of constituent quarks with structure.
For example, the proton is described in terms of three $UUD$ constituents with
an $SU(6)$ inspired wave function. At large $Q^2$ the virtual photon probes
deep into one constituent and sees its parton structure. The proton structure
function is obtained as a convolution of the $Q^2$ independent constituent wave
function with the $Q^2$ dependent constituent structure function. Similar
models of the nucleon in terms of constituents with structure have been
considered over the years also to describe the static properties of nucleons
\cite{dillon,hwa,pene}. In our case, the nucleon structure function is treated
in full analogy with the case of Helium 3, with constituent quarks replacing
nucleons. Of course one may object that nucleons, i.e. the constituents of
Helium-3, are colourless and therefore can exist as unconfined units. On the
contrary, the constituent quarks are confined, so that they cannot be really
independent of each other and a colour field string must connect them to each
other. However, it is conceivable that the content of the string in terms of
sea partons and gluons could be small in comparison with the structure of the
constituent. Alternatively, a string segment could be associated with the
constituent in a universal way, independent of the constituent flavour and of
the hadron, so that, in a sense, it becomes a part of the constituent itself.
At the other extreme, the string could be responsible for the whole structure
of the hadron. In this extreme case we would have a model of three
structureless valence quarks and a sea of quark and gluons from the string
\cite{kuti} in principle different for different hadrons. The real hadron will
probably be somewhat in between.

In this note we discuss the quantitative information  that can be obtained on
this issue from the available data on the pion structure function which have
been collected from measurements of the Drell--Yan lepton pair production
cross-section. In the proposed model where all the structure is in the
constituents, one can start from the known parton densities in the nucleon,
deconvolute the wave function and obtain the parton densities in the
constituents. From these one can then predict the pion structure function,
given a reasonable wave function for the pion. We will compare the predictions
of this model with the data on the pion structure function obtained from the
Drell--Yan process \cite{whalley}. This kind of comparison has already been
done long ago \cite{pene,rapuano} but the data on the pion structure function
\cite{sutton} are by now sufficiently precise to make the present re-analysis
worthwhile. In fact the validity of the constituent-with-structure ansatz can
now be significantly tested.~We shall see that the model is in reasonable
albeit not perfect agreement with the data.

It is true that the choice of the wave functions introduces some ambiguity in
the prediction of the pion structure function from that of the proton. But
there are important sum rules in this model that are valid independently of the
wave function form. In fact the amount of momentum carried by gluons, by sea
and by valence should be separately the same in the nucleon and in the pion at
the same $Q^2$ (as a consequence of the fact that the constituents carry the
totality of the hadron momentum). No such equality is predicted by the model
where all the sea and gluons, or a substantial part of them, arises from the
string, the structure of the string in the proton and in the pion being in
principle different. The separate determination of sea and gluons in the pion
is difficult, because the available Drell--Yan data do not give any information
on the pion structure functions at $x \leq 0.2$. Most of the information on the
gluon distribution in the pion arises from the limited data on large $p_T$
photons produced in $\pi^+p$ reactions \cite{sutton}. But the total momentum
carried by sea and gluons is well determined being the complement to 1 of that
of valence and one finds 0.61$\pm$  0.02  for the proton \cite{martin} and
0.54$\pm$0.04 for the pion \cite{sutton}, at $Q{_0^2} = 4$~GeV$^2$. The results
of this analysis appear to support to a fair degree of accuracy the
constituents-with-structure model.

 We recall that another application of the
formalism of Ref. \cite{acmp} is for nuclei. In Ref. \cite{zhu}, it is shown
that a substantial part of the EMC effect (the $A$ dependence of the nucleon
structure functions) can be attributed to the distortion of the constituent
wave function inside a nucleon due to the external nuclear field. The model can
also be applied to polarized deep inelastic scattering \cite{altrid}. In this
model the constituents carry the totality of the proton spin and the observed
{\it spin crisis} is described by a corresponding
depletion of the fraction of the constituent spin which is
carried by parton quarks. In this picture it is particularly clear that the
experimental results are not at variance with the constituent model. Rather
they have implications on the constituent structure.

\section{The model}

	For definiteness consider a proton $p$ or a positively charged pion
$\pi^+$. In the
model where the structure of the hadron is due to the structure of the
constituents, the
parton density $r_h(x,Q^2)$ for a given parton type $r$ in the hadron
$h$ is given by \cite{acmp}:
\beq
r_h(x,Q^2) =
\int^1_x \frac{dz}{z} \left[ U_h(z) r_U \left( \frac{x}{z},Q^2 \right) +
D_h(z) r_D
\left(\frac{x}{z},Q^2 \right) \right]~,
\label{eq:convol}
\eeq
where $U_h$ is
the density of up-constituents in the hadron $h = p, \pi^+, D_p$ is the
density of
down-constituents in the proton $p$, while $D_{\pi^+}$ is the density of
$\bar D$
(antidown) constituents in the pion $\pi^+, r_{U,D}$ are the parton
densities in the $U$
or $D$ constituents (for the pion $r_D$ is actually $r_{\bar D}$ which is
the same as
$\bar r_D$). As the $Q^2$ evolution matrix does not depend on the
target, i.e. it is
the same for the partons in a proton or in a constituent, it follows
that if the
convolution is valid at one $Q^2$ it will remain valid at all $Q^2$.
Note that the
moments
\beq
r^{(n)}_h(Q^2) = \int_{0}^{1}dx x^{n-1}r_h(x,Q^2)
\eeq
are simply given by a sum of products of
moments:
\beq
r^{(n)}_h(Q^2) = U^{(n)}_h r^{(n)}_U (Q^2) + D^{(n)}_h r^{(n)}_D(Q^2)~.
\eeq
For short hand we indicate the above convolution by
\beq
r_h = \left[ U_h \otimes r_U + D_h \otimes r_D \right]~.
\eeq
For example, the gluon density in the proton is given by
\beq
g_p = \left[ U_p \otimes g_U + D_p \otimes g_D \right] = \left( U_p +
D_p\right) \otimes
g_U
\eeq
where the last step is due to the equality of the gluon density in $U$
and $D$
constituents. Actually it is important to note that, by using obvious
isospin relations
(like $u_D = d_U$, etc.), for all partons kinds $r_p$
one can refer to the densities of
partons in the
$U$ constituent. This is also true in the pion case. Now recall that the
first moment
of $U$ and $D$ are $U^{(1)}_p = 2, D^{(1)}_p = 1$ while, for the second
moments, $U^{(2)}_p +
D^{(2)}_p = 1$,
because constituents carry the totality of charge and momentum of the
proton. Hence
$g^{(2)}_p = g^{(2)}_U$.  Clearly, for similar reasons, also
$g^{(2)}_\pi = g^{(2)}_U$.
Thus, independent of the wave functions, the total momentum of gluons in
$p$ and in
$\pi^+$ are predicted to be the same at the same $Q^2$. By an identical
argument, the same
prediction holds for the
total sea second moment and consequently for the total momentum
 carried by valence.

\section{Parameters of the model}

	In order to predict the parton densities in the pion from those in the
proton, we take
the proton parton densities given by the most recent fits of all
available data obtained
by Martin et al. in Ref. \cite{martin}.   Precisely we use the set of parton
densities labeled by
MRS(G), with the two-loop $Q^2$ evolution evaluated for $\Lambda =
255$ MeV,
where $\Lambda$ refers to $N_f = 4$ in the $\overline{MS}$ definition,
corresponding to $\alpha_s(m_Z) = 0.114$. Other available sets of
structure functions will
be used to check the stability of the results (also with different
values of $\Lambda$).
As for the distributions of the $U$ and $D$ constituents in the proton
we take those given
in Ref. \cite{acmp}. These constituents densities, based on SU(6)$_W
\otimes$ O(3) at
$p_z \rightarrow \infty$, are complicated and will not be reproduced
here (it suffices to say that the parameters introduced in Ref. \cite{acmp} are
fixed to the values $\beta=0.44$, $a^2=0.8$). Simpler choices
would also work and we have checked that no important changes in the
final results are
obtained with different starting wave functions. Then for the parton
densities in the $U$
constituent we adopt the following parametrisation at $Q{_0^2}=4$~GeV$^2$

\bea
q_U(x,Q_0^2)&=&
C_s \frac{(1-x)^{(D_s-1)}}{x^{(1+d_s)}} +
\delta_{uq} \frac {B^{-1}(A,\frac{1}{2}) (1-x)^{(A-1)}}{\sqrt x} ,
\nonumber \\
\\ \label{eq:parton}
g_U(x,Q_0^2)&=&C_g \frac {(1-x)^{(D_g-1)}}{x^{(1+d_g)}} .
\nonumber
\eea

Where B(x,y) is the Euler beta function. In the quark formula there is a
valence term, only present for the u parton quark, and a universal sea term.
The differences in the sea composition  (strange vs. non-strange, $\bar u$ vs.
$\bar d$  etc.)  are irrelevant here and have been neglected. At small $x$, a
stronger behaviour than $1/x$ for sea and gluon densities, parametrised by the
positive coefficients $d_s$ and $d_g$, has been allowed according to the
results obtained by ZEUS and H1 at HERA. Of course the  momentum sum rule
imposes a relation among the parameters.

The parameters appearing in the above formulae were fitted to reproduce the
input parton densities in the proton, given the chosen wave function. The
values of the parameters obtained from the fit are: $A=0.776$, $C_s=0.5$,
$D_s=3.3$, $d_s=0.085$, $D_g=1.3$, $\delta_g=0.45$. The comparison between the
input parton distributions of Ref. \cite{martin} at $Q{_0^2}$ and the results
of the fitted distributions for the model are shown in Fig.~1. As seen, given
the accuracy to which the parton densities are known, a very good fit is
obtained. Thus there is no doubt that the proton data are nicely consistent
with the model.

\begin{figure}[t]   
    \begin{center}
       \setlength{\unitlength}{1truecm}
       \begin{picture}(9.0,9.0)
         \put(-7.0,-10.0)
         {\includegraphics{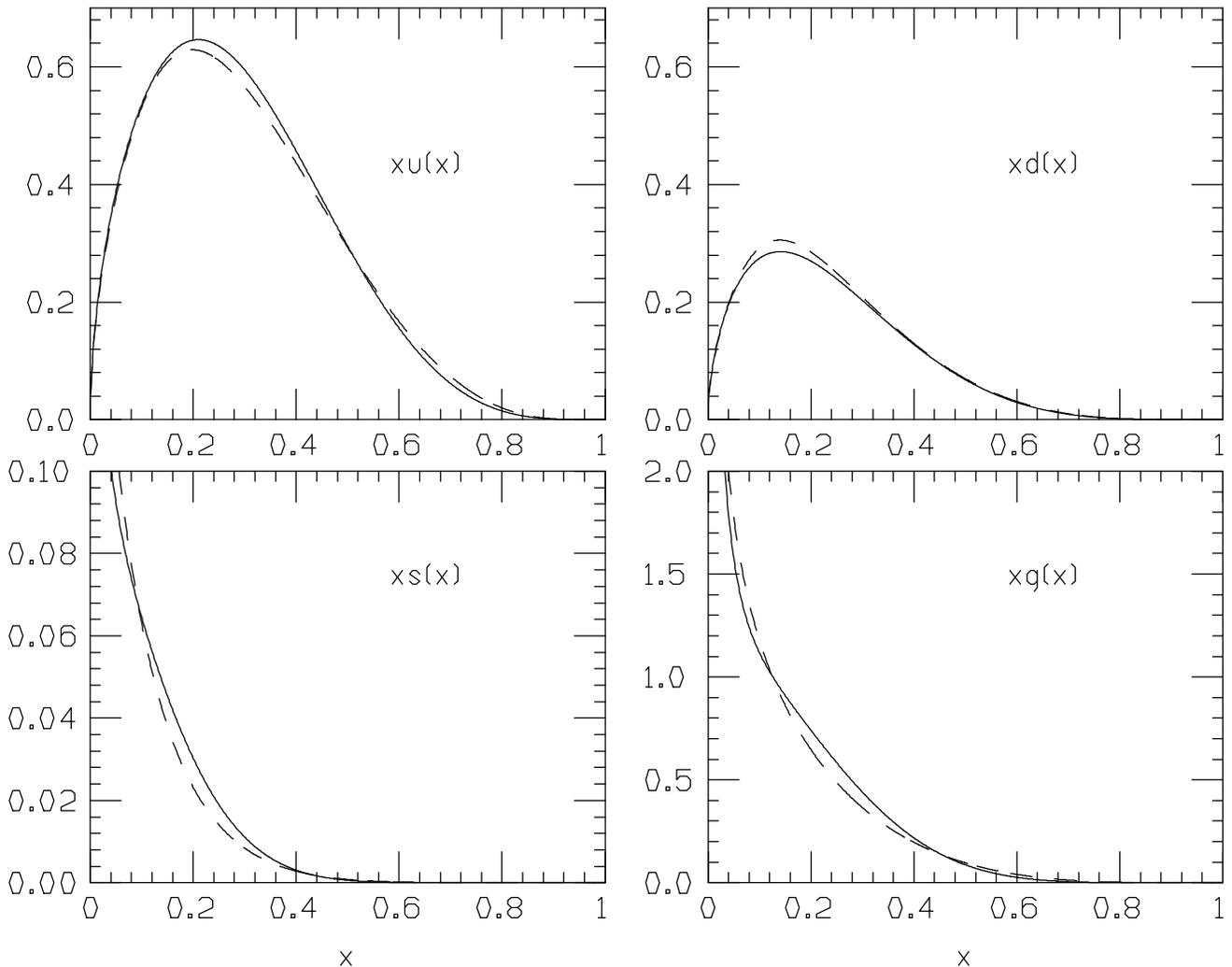}}
       \end{picture}
    \end{center}
       \vskip 3cm
\caption{Proton parton densities at $Q^{2}_0=4$~GeV$^2$. Solid lines refer to
the fit of MRS(G) \protect\cite{martin}; dashed lines refer to the present
model. The differences are well within the uncertainties of the MRS(G) fit.}
\protect\label{fig.1}
\end{figure}

For the total momentum fraction carried by valence in the proton at $Q{_0^2}$
one has:
\beq
V^{(2)}_p (Q{_0^2}) = 0.39 \pm 0.02
\label{eq:momentum}
\eeq
where $V = [u - \bar u + d - \bar d]$. The error has been estimated by
reevaluating the moment starting from the available recent compilations
of the proton
parton densities, as shown in Table {\ref{tab1}}. We also varied the value of
$\Lambda$ in a range corresponding to $0.110 \leq \alpha_s(m_Z) \leq 0.125$
using the recent results of Ref. \cite{stir}. The total error is a combination
of the uncertainty at fixed $\alpha_s$ with that from varying $\alpha_s$.
The difference in $V_p^{(2)}$
using either the input nucleon densities or the
fitted densities in
the $U$ constituent is completely negligible given the quoted error.

\begin{table}
\centering
\begin{tabular}{|c|c|c|c|}\hline
$$ &$\alpha_s(m_Z)$& $v^{(2)}_p (Q{_0^2})$ & $g^{(2)}_p (Q{_0^2})$\\
\hline
CTEQ1M   &  0.111 & 0.390  &  0.419   \\
MRSS0    &  0.110 & 0.386  &  0.448   \\
MRSD-    &  0.110 & 0.383  &  0.444   \\
MRS(G)   &  0.114 & 0.392  &  0.427   \\
MRS 110  &  0.110 & 0.377  &  0.434   \\
MRS 115  &  0.115 & 0.382  &  0.428   \\
MRS 120  &  0.120 & 0.390  &  0.421   \\
MRS 125  &  0.125 & 0.398  &  0.414   \\
\hline
\end{tabular}
\caption{Second moment of valence, $v^{(2)}_p$, and gluon, $g^{(2)}_p$, in
the proton at $Q{_0^2}=4$~GeV$^2$ for some of the most recent parton density
 parametrizations.}
\label{tab1}
\end{table}

\section{Results}

	Having derived the parton densities in the $U$ constituent we now
proceed to predict the
pion structure functions. The only ingredient which is still needed is
the distribution
of $U$ and $\bar D$ constituents in the pion. Following Ref. \cite{rapuano}
 we take
\bea
U_{\pi^+}&=&{\bar D}_{\pi^+}=1/2 V_{\pi^+},\nonumber \\
\\ \label{eq:consti}
V_{\pi^+}(x)&=&\sqrt{ \frac {8 \tilde \beta} {\pi} } \frac {1} {x(1-x)}
         exp{[-2 \tilde \beta ln^2 \frac{x}{1-x}]},
\nonumber
\eea
where the parameter $\tilde\beta$, is fixed to the value $\tilde\beta$=0.1 in
such a way as to approximately have
$x \cdot v_{\pi}(x,Q{_0^2})\sim(1-x)$ as $x\rightarrow1$,
according to the Drell--Yan--West relation \cite{dyw}, and $v_{\pi}$ is the
valence quark distribution in the pion.

The predictions for the parton densities in the pion at $Q_0^2$ are simply
obtained by convoluting according to eqs. \ref{eq:convol}, the constituent
distributions in eq. 8 with the parton densities in the
constituent specified in eqs. 6 as a result of the proton fit.
In particular, as already mentioned, the predicted
second moment of valence coincides with the result
for the proton and is given in eq. \ref{eq:momentum}.

The above predictions should now be confronted with the experimental data from
Drell--Yan processes \cite{whalley}. The way to extract the pion structure
functions from the Drell--Yan data has been recently discussed in Ref.
\cite{sutton}. As it is well known the Drell--Yan cross-section is obtained by
a convolution of the parton densities in the proton times those in the pion
times the partonic cross-section. The latter includes the QCD correction which
leads to a quite substantial $K$ factor \cite{ellis} at the relevant dimuon
mass scale (typically between the $J/\psi$  and the $\Upsilon$). Thus in
principle in order to extract the pion densities one has to compute the $K$
factor. The authors of Ref. \cite{sutton} chose to write the $K$ function in
the form $K(x_F,Q^2) = K^{(1)} K^\prime$, where  $K^{(1)}$ is the simple $K$
factor computed at one loop accuracy, while $K^\prime$ includes the effect of
higher orders (including the correction for a possible bad choice of $\alpha_s$
in the leading term). In Ref. \cite{sutton} $K^\prime$ was fitted from the
data. This procedure is only justified if $K^\prime$ is really a constant in
$x_F$. Indeed the fit is sensitive to a constant rescaling of $V$. In fact the
valence is the dominant contribution at the rather large values of $x$ where
the data for the pion structure function exist and the overall scale of valence
is normalized by its first moment $V^{(1)}_\pi = 2$, where $V_\pi = u+\bar d$.
A consistency check is that $K^\prime$, arising from higher orders, should come
out reasonably close to 1. In the fits of Refs. \cite{sutton} $K^\prime$ ends
up in a range between $1.1 \div 1.3$.

If indeed $K^\prime$ is with good approximation a constant, we can take the
results of the fits in Ref. \cite{sutton} as a compact description of the
actual data. Recently an almost complete calculation of the rapidity dependence
of the $K$ factor at two loop accuracy has been performed in Ref.
\cite{vannee}. This calculation is not complete because the effect of soft
gluon contributions (within a specified definition) is not included. We have
repeated the procedure of Ref. \cite{sutton} with the available two-loop QCD
corrections to the Drell--Yan $x_F$ differential cross-section. We found that
with a very good accuracy $K^\prime$ is indeed a constant over the rapidity
range of the experiment. Thus we can validate the procedure of Ref.
\cite{sutton}.  Of course further uncertainties on the result of the fit beyond
the statistical accuracy arise from the rudimentary parametrization adopted for
the pion densities and from the assumptions made for the sea and gluon
densities which the data do not much constrain.

In Fig.~2 we present a comparison between the model fit of the data and the
best fit obtained in Ref. \cite{sutton}.  The theoretical predictions are
presented with and without inclusion of the  $K^\prime$ factor. We see that the
model fits the data quite well but at the price of a somewhat larger
$K^\prime$. While in the model independent fit $K^\prime$ ranges between 1.1
and 1.3, in the model one needs a larger $K^\prime$, $K^\prime  = 1.3 \div
1.6$.

\begin{figure}[t]   
    \begin{center}
       \setlength{\unitlength}{1truecm}
       \begin{picture}(9.0,9.0)
         \put(-7.0,-10.0)
         {\includegraphics{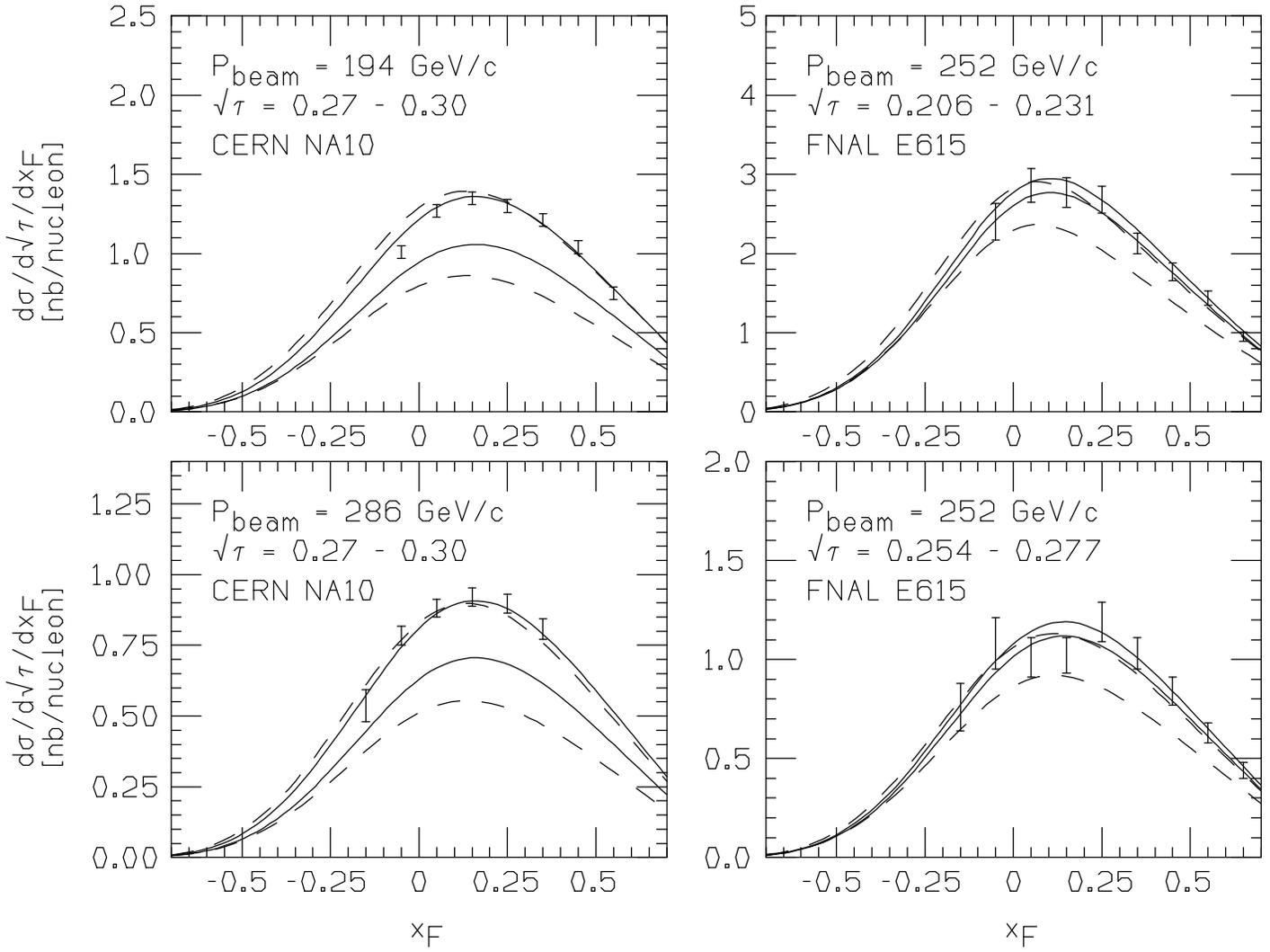}}
       \end{picture}
    \end{center}
       \vskip 3cm
\caption{A comparison of a sample of experimental Drell--Yan data
\protect($\pi{^{-}}$W reactions) with the fit in Ref. \protect\cite{sutton}
(solid lines) and with the present model (dashed lines). In each case we
present result with (upper curves) and without (lower curves) the K' factor.
Thus both sets of curves can fit the data rather well but the model needs a
larger K' factor.}
\protect\label{fig.2}
\end{figure}

Clearly the model independent fit has a slightly better $\chi^2$ than the
model.
Also, in the model, the resulting values of the $K^\prime$ factor are a bit too
large to be really satisfactory.
This difference of $K^\prime$ factors is a consequence of the different values
for the second moment of valence found in the model independent fit and for
that implied by the model:
\bea
V^{(2)}_\pi (4~{\rm GeV}^2) & = & 0.46 \pm 0.04~{\rm (fit)} \\
V^{(2)}_\pi (4~{\rm GeV}^2) & = & 0.39 \pm 0.02~{\rm (model)}
\eea
where, of course, the model value is the
same as in the proton. The error attributed to the fit is our estimate which
takes into account the error within the procedure of Ref. \cite{sutton}, as
given by the
authors, plus the ambiguities related to the assumptions made in the procedure
(mainly from the parametrisation choice for the pion, the $x_F$ independence of
$K^\prime$, the value of $\alpha_s$ etc.).
A plot of the resulting structure function of the pion, in the present model,
is shown in Fig.~3, where it is compared with the fit of \cite{sutton}.

\begin{figure}[t]   
    \begin{center}
       \setlength{\unitlength}{1truecm}
       \begin{picture}(9.0,9.0)
         \put(-7.0,-10.0)
         {\includegraphics{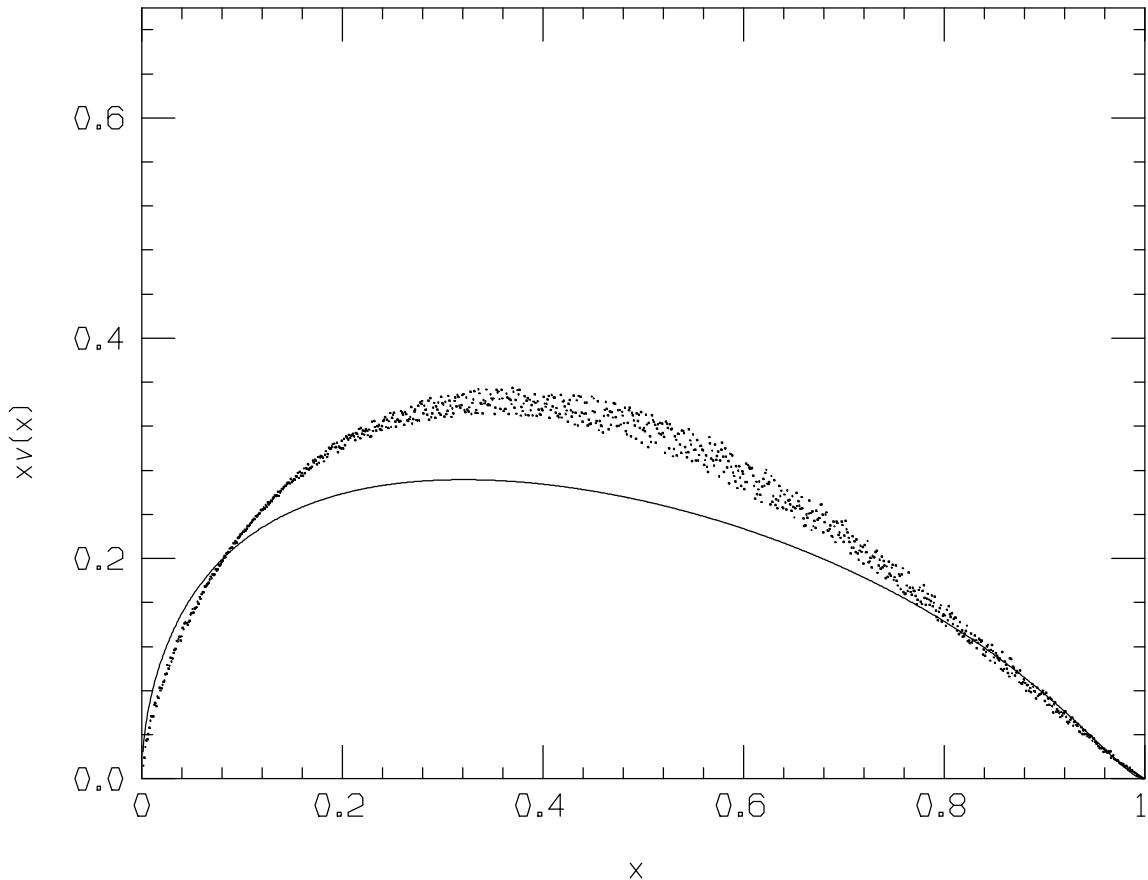}}
       \end{picture}
    \end{center}
       \vskip 3cm
\caption{The resulting valence distribution $x \protect\cdot v_\protect\pi(x)$
in the model (solid line) compared with the fit of Ref. \protect\cite{sutton}
given as a band that includes the uncertainties on their procedure.}
\protect\label{fig.3}
\end{figure}

\section{Conclusions}

In conclusion, the constituent-with-structure model is shown to provide a
reasonably accurate description of the pion structure functions as determined
by experiment. This is particularly remarkable in that the pion is a very
peculiar hadron with mass that vanishes in the chiral limit. Thus there  is a
strong indication that the model can actually provide a reasonable first
approximation of the structure functions of any other hadron for which no data
exist.

Finally, we recall that the second and third moments of the valence parton
densities in the pion have been estimated in lattice QCD in the quenched
approximation
\cite{sachra}.
There the result for the second moment was $V^{(2)}_\pi(49~{\rm GeV}^2) = 0.46
\pm 0.07$ which corresponds to $V^{(2)}_\pi(4~{\rm GeV}^2) = 0.55 \pm 0.08$.
This is a rather large value in comparison with the prediction of the model.
However, it is known that quenched lattice calculations fail to reproduce the
momentum fractions carried by up and down quarks in the proton \cite{schi}.
For the proton the lattice results are larger then the fitted values by at
least a factor of two.
Thus the conclusion is that the quenched approximation appears to be
rather poor for the calculation of hadronic structure functions.

\bigskip
We warmly thank W.J.~Stirling for providing us the fortran code of the MRS(A')
and MRS(G) parton density parametrizations, P.J.~Rijken and W.L.~van~Neerven
for the two loop Drell--Yan cross section program and G.~Martinelli for the two
loop $Q^2$ evolution program.

\end{document}